\documentclass[manuscript]{aastex}

\slugcomment{Based on observations obtained with Asiago 1.82 m telescope}

\shorttitle{A high resolution spectral atlas of peculiar stars}
\shortauthors{Tomasella et al.}

\begin{document}

\title{A high resolution, multi-epoch spectral atlas of peculiar stars 
    including RAVE, GAIA and HERMES wavelength ranges}

\author{Lina Tomasella\altaffilmark{} and Ulisse Munari\altaffilmark{}}
\affil{INAF Osservatorio Astronomico di Padova, 36012 Asiago (VI), Italy}

\author{Toma\v{z} Zwitter\altaffilmark{}}
\affil{Faculty of Mathematics and Physics, University of Ljubljana, Ljubljana, Slovenia}
\affil{Center of excellence SPACE-SI, Ljubljana, Slovenia}

\begin{abstract}
We present an Echelle+CCD, high S/N, high resolution ($R$ = 20\,000)
spectroscopic atlas of 108 well-known objects representative of the most
common types of peculiar and variable stars.  The wavelength interval
extends from 4600 to 9400~\AA, and includes the RAVE, Gaia and HERMES
wavelength ranges.  Multi-epoch spectra are provided for the majority of
observed stars.  A total of 425 spectra of peculiar stars are presented,
which have been collected during 56 observing nights between November 1998 
and August 2002.  The spectra are given in FITS format and heliocentric wavelengths,
with accurate subtraction of both the sky background and the scattered
light.  Auxiliary material useful for custom applications (telluric
dividers, spectro-photometric stars, flat-field tracings) is also provided. 
The atlas aims to provide a homogeneous database of the spectral appearance
of stellar peculiarities, a tool useful both for classification purposes and
inter-comparison studies.  It could also serve the planning for and training
of automated classification algorithms designed for RAVE, Gaia, HERMES and
other large scale spectral surveys. The spectrum of XX~Oph is discussed
in some detail as an example of the content of the present atlas.
\end{abstract}


\section{Introduction}

Stars are considered peculiar if the observable
characteristics place them aside from the bulk of normal stars, especially
those in quiet long-lived phases of stellar evolution. The
boundaries are vaguely defined, even if the presence of pulsations (like in
Cepheids, RR Lyrae, Miras, etc.), chemical anomalies (Ba, Am, Carbon, CH
stars, etc.), active chromospheres (BY Dra, flare stars, RS CVn, etc.),
emission lines (Be, Wolf-Rayet, etc.), mass transfer (cataclysmic variables,
$\beta$ Lyrae, etc.), outbursts (novae, symbiotics, etc.) and ionized
massive winds (LBVs, post-AGB, etc.) are unanimously considered as landmarks 
of peculiarity (e.g., Tout \& Van Hamme 2002).

Spectra of many peculiar objects have been published already, but they
are scattered throughout the literature, most papers dealing with
a single object at a time. This varied dataset is inhomogeneous in
terms of spectral resolution and wavelength range, spectrograph and 
detector type and in electronic availability of the spectra.  
Some extensive spectroscopic atlases contain also spectra 
of peculiar stars (e.g., Danks \& Dennefeld 1994) or are entirely 
devoted to some class of peculiar objects (e.g.,\ Allen 1984,
Barnbaum 1994), but they too come in a huge variety of formats and content. 
A recent and fairly complete census of ultraviolet/optical/infrared
spectroscopic databases was compiled by Sordo and Munari (2006), the {\it
Asiago Database of Spectroscopic Databases}\footnote{web query interface
http://web.oapd.inaf.it/adsd/} (ADSD).  ADSD surveyed 294 spectral 
libraries by comparing their main observational characteristics (like 
wavelength range, resolving power, fluxing, detector, data access and 
retrieval possibilities, etc.). It plotted typical spectra for a 
quick-look inter-comparison, and counted the included stellar types 
in terms of spectral type, luminosity class, peculiarity flags, and 
atmospheric parameters $T_{\rm eff}$, $\log g$ and [M/H].

Literature apparently lacks a homogeneous spectral atlas comprised of 
high-resolution, high S/N and wide wavelength range spectra and 
covering as diverse range of stellar peculiarities as possible. 
This is particularly relevant in view of the current and forthcoming 
massive all-sky surveys obtaining medium to high-resolution spectra, 
like RAVE, Gaia and HERMES. These surveys will collect spectra of 
millions of stars, extracting their radial velocities, atmospheric
parameters and chemical abundances. They will provide fundamental
results for Galactic Archeology, or how our Galaxy formed and evolved
(Freeman and Bland-Hawthorn 2002). These and other surveys observe 
over pre-defined and limited wavelength ranges, but the manifestation 
of spectral peculiarities is well known to change with wavelength.  
Thus, an atlas of peculiar objects extending over a broad range of 
optical wavelengths could assist in comparing the characterization 
of peculiarity from limited wavelength intervals to that gathered 
from broader ones. The survey we present here includes the wavelength 
ranges of all these spectroscopic surveys and is somewhat better in 
terms of spectral resolution and S/N. So it is useful for training 
of the surveys' automated classification algorithms.  

The aim of this paper is to provide such a homogeneous, high-resolution, high
S/N, and wide wavelength range atlas of a broad sample of peculiar stellar
objects.  We became interested in such a project when we were preparing the
spectral atlases of normal stars (Munari \& Tomasella 1999; Marrese, Boschi
\& Munari 2003) to be used during preparations for the Gaia mission, 
soon to be launched by ESA. Preliminary results on peculiar stars were 
reported by Munari (2002, 2003) and presented sample spectra over the Gaia 
and RAVE wavelength range, which is centered on the far-red Ca~II triplet 
and the head of the Paschen series of hydrogen between 8450 and 8750~\AA.
In the next two sections we discuss the target selection, observations and 
data reduction. A description of atlas products (Sec.~4) is followed by 
discussion of some examples of spectra included in the atlas. 

\section{Target selection}

We observed 108 well-known objects, sparsely selected from existing
literature, to document the main types of stellar peculiarity.  The program
objects are listed in Table~1, where they are grouped into six broad groups:
chemically peculiar; pulsating; interacting and eruptive binaries; stars
with active surfaces, rapid rotation; emission-line objects; and others. 
The columns of Table~1 give the (1-2) star name and its HD number, (3) type
of peculiarity, (4) type of variability from the living
edition\footnote{http://www.sai.msu.su/groups/cluster/gcvs/gcvs/} of the
General Catalogue of Variable Stars (GCVS), (5-6) spectral type and its
source, (7-8) the $(B-V)_T$ color and the $V_T$ magnitude as given in the
Tycho-2 catalog.  If the latter are not available, a typical $V$ magnitude
from other sources is given.

\section{Observations and data reduction}

The observations were carried out with the Echelle spectrograph mounted at
the Cassegrain focus of the 1.82 m telescope operated in Asiago by INAF
Astronomical Observatory of Padova.  The detector was a Thomson THX31156 CCD
with 1024 $\times$ 1024 pixels of 19 $\micron$ size cooled with liquid
nitrogen.  The chip was of the thick type, front illuminated, with Lumigen
coating for enhanced blue response and presented no detectable fringing in
the red.  The cross-disperser was a grating, thus the reddest Echelle orders
were contaminated by the superimposed second order from the cross-disperser. 
This contamination was suppressed with a high-pass OG455 filter with a
thickness of 3~mm which was inserted into the optical path.  The filter cut
the light with wavelengths bluer than 4600~\AA.  So it effectively removed
the second order contamination up to 9200~\AA, but it also set the short
wavelength limit of our atlas.  Great care was taken to keep the dispersion
and the resolving power values constant during the whole observing campaign
which started in November 1998 and was completed in August 2002, for a total
of 56 different observing nights.  All observations were carried out with
the spectrograph slit opened to a width of 2~arcsec.  The slit was aligned
with the parallactic angle when the airmass exceeded 1.5.  At that airmass,
the atmospheric differential refraction at the two ends of the recorded
wavelength range reaches $\sim$1.0~arcsec (cf.\ Filippenko 1982).

Our Echelle spectra cover the spectral range $\lambda\lambda$ 4600$-$9400
\AA\ in 25 orders.  The 15 bluest orders ($\lambda\lambda$ 4600$-$6890 \AA)
are characterized by a continuous spectral overlap, while inter-order gaps are
present for the redder orders. The following spectral windows
are not covered by our spectra: 6890-6896, 7104-7118, 7334-7356, 7578-7610,
7840-7882, 8120-8173, 8421-8488, 8745-8828, and 9095-9196~\AA.

The instrumental PSF (measured from the FWHM of unblended telluric absorption
lines of O$_2$ and H$_2$O) remained pretty constant during the observing
period, corresponding to a resolving power $R$ = 20\,000.  The PSF,
deduced by comparing the FWHM of the unblended lines in the Thorium 
wavelength calibration spectra, was found to be generally uniform 
over various Echelle orders.

The height of the slit on the sky was 10 arcsec. The program star 
was always placed close to one end of the slit. The spectrum extracted 
from the other half of the slit height was not illuminated by the star.
So it was used to derive an accurate median sky spectrum which was subtracted 
from the stellar tracing. Most of the program stars were quite bright objects 
which required short exposure times, so the background sky contribution 
was very weak.

All the data reduction and calibration was carried out in IRAF.  Spectral
tracing was performed by weighting the extraction according to the variance
of the recorded spectrum.  The wavelength calibration was quite accurate,
with a typical rms of 0.3 km/s for all program stars.  Scattered light was
carefully modeled and subtracted from the bi-dimensional frames. All spectra 
were inspected for possible residuals of the strongest
night-sky emission lines (principally [OI] 5577, 6300, 6364~\AA). This 
confirmed the accuracy of sky subtraction.

Each final spectrum presented in this atlas was built starting from several
(at least three) individual exposures obtained consecutively at the
telescope.  Each individual exposure was extracted, reduced and calibrated
separately, before summing them into the final spectrum. Before merging, 
the resulting individual spectra were carefully compared
order by order to spot and remove the presence of cosmic rays or other
defects.  We restrained from doing this automatically during data reduction, 
principally to avoid potential problems with very sharp emission lines 
and to keep a strict control over the whole data reduction process.

Cool objects could have their reddest Echelle orders saturated while the 
blue part of the spectrum would remain under-exposed. In such cases we 
obtained consecutive observations with varying exposure times. The final 
spectrum was obtained by combining the best exposures of individual 
Echelle orders and scaling their counts to a uniform exposure time.
Multiple exposures with varying exposure time were also used for objects
with bright emission lines that saturated the deep exposures. Shorter 
exposures were used to recover the unsaturated emission line profiles, 
usually those of the H$\alpha$ line.

The procedure described above gave the final spectrum of a given object 
at one observing epoch. But the majority of the program stars in this atlas
were observed at more than one observing epoch. For objects that were
not expected to vary significantly in time (like for example the chemically
peculiar stars), these multi-epoch spectra were generally obtained within
the same observing night.  The intention is to provide a larger set of
independent spectra which can be used to measure e.g.\ the equivalent widths 
or to derive profiles of spectral lines.  For objects with a large variability
the multi-epoch spectra have been generally collected during different
nights.

\section{Atlas products}

The 425 spectra of the 108 program stars presented in this atlas are
available as FITS files from the CDS\footnote{http://cdsweb.u-strasbg.fr}. 
Their wavelength scales are heliocentric.  The ordinates of the spectra
correspond to photo-electrons (i.e.\ counts multiplied by the ADU conversion
factor, so that the local S/N can be estimated directly from the
spectrum).  Table~2 provides a complete list of the spectra
made available in this atlas.  Its columns give: (1) the root of the FITS
file name, (2) the heliocentric JD at mid-exposure, (3) the total exposure
time (in seconds of time), (4-5-6-7) the name and HD number of the star, the
type of peculiarity and its spectral classification (all four repeated from
Table~1), (8) spectral range covered by the spectrum, (9) corresponding
Echelle orders, (10) the S/N on the continuum adjacent to H$\alpha$, (11-12)
related spectrophotometric standard(s) and telluric divider(s) numbered
according to Tables~3 and 4 (see below), and, finally, (13) some notes.

We did not correct for the blaze function of the Echelle orders and thus we
left the spectra in their native multi-order format without merging them
into single-dispersion, mono-dimensional outputs. We left to correct it to the
motivated user of this atlas according to specific needs. The reason for this
is that correcting for the blaze function is a procedure that can be carried 
out in different ways and it is never trouble-free, for the following
reasons.

The removal of the blaze function could be performed by fluxing the spectra
against suitable spectro-photometric standard stars.  To ensure successful
results, the absolute flux of the standard stars should be accurately and
continuously sampled at high spatial frequency (e.g.\ in 1~\AA\ steps, given
the limited wavelength extension of a single order), the observations should
be carried out under all-sky photometric conditions, and the position on the
spectrograph slit should be the same for the standard and the program stars
(so to ensure similar illumination of the optical train within the
spectrograph).  At the time we carried out the observations described in
this paper, standard stars accurately calibrated at 1~\AA\ steps did not
exist, and the sky conditions were not always photometric.

The shape of the blaze function could in principle be estimated also from
flat-field exposures.  Unfortunately, the color temperature of the
flat-field lamps is generally quite different from that of the observed
stars, and in addition the even illumination of the slit by the flat-field
differs significantly from that produced by a star.  While using flat-field
exposures could remove most of the blaze function, the results are never
accurate enough to allow a smooth joining of the adjacent echelle orders.

Finally, it could also be possible to get rid of the blaze function by
continuum normalization of the individual echelle orders.  This is a procedure
generally carried out by a trial-and-error application of interpolation
functions (spline, polynomials, etc.). It requires a human intervention and
guessing about the {\it true} shape of the underlying stellar continuum. 
The latter is frequently compromised by the extended wings of emission lines,
molecular absorption bands, lack of useful stretches of unpolluted continuum 
(for example because there are too many absorption and/or emission lines), 
and even worse by the huge breadth of the Balmer lines in early type stars, 
that can encompass a whole echelle order.

In addition to stellar spectra, we provide also other products: flat field
tracings, spectro-photometric standards, and telluric dividers. They are 
described in the rest of this section.

Flat fields were exposed on a dome white-screen, uniformly illuminated by a
3750~K halogen lamp.  Flat field tracings were extracted for each stellar
spectrum presented in this atlas, by adopting exactly the same tracing and
weighting parameters.  The flats are named using the same root name as the
corresponding stellar spectrum (for ex.  AFDra$\_$1.fits is the stellar
spectrum, and flat$\_$AFDra$\_$1.fits is the corresponding flat).  The flat
fields presented in this atlas have a very high S/N (frequently in excess of
300).  The {\it noise} observed in the flat fields is well accounted by the
sole Poissonian statistics of the exposure level, with pixel-to-pixel
differences being negligible contributors.  This confirms that the CCD chip
we used had an excellent cosmetics and was not affected by systematics like
fringing in the red.  We provide the flat fields for each star mainly (1) to
allow interested user to experiment with correction of the blaze function,
and (2) to check for the very rare appearance of the feeble shadow a dust
grain laying on the entrance window of the CCD dewar.  It is worth noticing
that at the reddest wavelengths, the absorption by O$_2$ and H$_2$O are so
strong that even the few meters traveled by the light within the telescope
dome are enough to mark the flat fields with absorption lines.

As remarked above, at the time the observations described in this atlas were
performed, no suitable spectro-photometric standard stars were available to
efficiently support Echelle observations.  In fact, Echelle spectra
presented at the time in literature were never fluxed into absolute or
relative fluxes, a practice rarely attempted even now.  Nevertheless,
observations of some spectro-photometric standards, selected from Hamuy et
al.  (1992, 1994) and Burnashev (1985), were obtained with the same
instrumental set-up and reduced in the same way as for the program stars. 
The spectrophotometric standards were observed irregularly during the
observing campaign, and the available ones are summarized in Table~3.  In
Table~2, we list in column 11 the spectrum of the spectrophotometric
standard stars that were observed during the same observing run of the given
program star.  The respective air-masses are available in Table~3.

The O$_2$ and H$_2$O in the Earth's atmosphere produce sharp absorption lines
on recorded stellar spectra. Their intensity depends on airmass and water
vapor content in the Earth's atmosphere at the time and along the
line of sight of the observation. These telluric absorption lines are best
visible against the featureless continuum of early type stars affected by
a high projected rotational velocity, that we name telluric dividers.
During most of the observing runs we obtained a very high S/N spectrum of  
at least one such telluric divider. Their spectra, extracted and treated
exactly as those of the program stars, are made available (together with 
the corresponding flat fields) with this atlas, and are listed in Table~4.
Column 12 of Table~2 reports the telluric divider(s) observed during the
same observing run of the given program star. The interested reader can use
these telluric dividers to compensate for the O$_2$ and H$_2$O absorption
lines present on the science spectra at the longer wavelengths. 

We did not correct ourselves the spectra for telluric absorptions for
basically three reasons.  The removal of telluric lines requires spectra
corrected for the blaze function.  An accurate removal, even over a short
wavelength range, is a trial-and-error procedure that requires human
intervention.  Finally, the intensity of the O$_2$ and H$_2$O telluric lines
ranges from optically thin, to optically tick, to completely saturated.  The
latter cannot be compensated for.  Optically tick lines require a lot of
devoted work and their successful removal is in not guaranteed.  Only
optically thin lines are easy to deal with.

\section{An example of the spectra included in the atlas}

For sake of providing a graphical guidance to the content of this atlas,
Figures~1 and 2 present some sample portions of the spectrum of the peculiar
star XX~Oph.  The spectrum is presented with the continuum normalized to
1.0. The major emission and absorption lines are identified.

XX Oph was first noted by Merrill (1924) that called it "the iron star" for
the many bright emission lines of ionized iron and metals, that were later
catalogued by Merrill (1951) and Cool et al.  (2005).  The optical spectra
of XX Oph are characterized by a hot continuum emission from nebular
material at blue wavelengths, from which molecular absorption bands of an
M6 III star emerge at red wavelengths (de Winter \& Th\'{e} 1990).  Spectral
variability of XX~Oph was studied by Goswami, Rao and Lambert (2001), and by
Tarasov (2006) among others.

XX Oph has an unusual photometric variability. Using Harvard collection of
photographic plates, Prager (1940) showed that the light curve of the star
had over 1-mag deep aperiodic minima in the blue-photographic region, which
could last for several years and are reminiscent of obscurations in R CrB
type of stars.  The 1964-2010 visual light-curve collected by AAVSO shows a
much lower degree of variability, with just two minima occurring in 1967
and 2004 superimposed to a steady linear decline from $m_{\rm vis}$=8.7 to
$m_{\rm vis}$=9.2 mag.  No outburst has been recorded over the 1890-2010
monitored period.

The spectrum of XX~Oph presented in Figure~1 shows the highly popular
regions of H$\alpha$, He/NaI and multiplets 42 and 49 of FeII.  The emission
lines are dominated by metals in their first stage of ionization.  They are
very sharp and their heliocentric radial velocities are close to $-$37
km~s$^{-1}$.  Only Balmer lines present a complex, non-symmetric profile
with a radial velocity of $-$18 km~s$^{-1}$ for the peak of the emission. 
Broad and blue shifted absorptions are visible for the strongest permitted
lines of FeII, CrII, TiII and NaI at radial velocities around $-$360
km~s$^{-1}$.  Table~5 summarizes the mean velocities for the photocenters of
the emission and absorption lines identified in Figure~1, which are similar
to the values reported earlier by Merrill (1961). The values given for 
H$\alpha$ and H$\beta$ refer to the velocity of the sharp peak in their
broad profile.

The H$\alpha$ profile of Figure~1 has no counterpart among those so far
published, with the exception of the H$\beta$ profile for July 1996
presented by Goswami et al.  (2001) for which the terminal velocity of the
H$\beta$ absorption component was $-$390 km~s$^{-1}$, while in our spectrum
it is $-$530 km~s$^{-1}$. The NaI profile presented in Figure~1 has never been
seen before in XX~Oph. While the sharp emissions at $-$37
km~s$^{-1}$ and the interstellar components at $-$11 km~s$^{-1}$ were
already known, the broad components at $-$365 km~s$^{-1}$ are entirely
new (cf.\ Tarasov 2006, Goswami et al.\ 2001, Merrill 1951).

There is a large uncertainty about the reddening affecting XX Oph. Lockwood
et al.  (1975) derived $E_{B-V}$$\approx$1.3, while Evans et al.  (1993)
preferred $E_{B-V}$$\approx$0.5.  The KI 7699 \AA\ interstellar line at
$-$11 km~s$^{-1}$ has an equivalent width of 0.186 $\pm$0.007 \AA\ on our
spectrum.  Using the calibration by Munari and Zwitter (1997), this
translates into a reddening $E_{B-V}$=0.73 $\pm$0.03.  This value agrees
with the intensity of the Diffuse Interstellar Bands at 5780, 5797, 5850,
6196, 6203 and 6614 \AA\ that are visible on our spectrum and with the
core-saturated interstellar components of NaI in Figure~1.

The top panel of Figure~2 shows the spectrum of XX~Oph over the wavelength
range covered by the ongoing RAVE survey of the southern sky (Steinmetz et
al.  2006, Zwitter et al.  2008) and the coming ESA's GAIA space mission
(Munari 2003).  The same wavelength interval is also tentatively base-lined
for a later phase of the LAMOST
survey$\footnote{http://www.lamost.org/website/en}$.  The bottom panel of
Figure~2 shows the spectrum of XX~Oph over the bluest (4708--4893 \AA) of
the four wavelength ranges to be covered by the forthcoming HERMES
all-sky survey$\footnote{http://www.aao.gov.au/AAO/HERMES/}$ 
(Freeman et al.\ 2010). The other three ranges are 
5649--5873, 6481--6739, and 7590--7890~\AA, all covered 
by the present atlas.

\acknowledgments

We are grateful to Federico Boschi and Paola Marrese for securing some
spectra included in this atlas.

\clearpage
\begin{figure}
\includegraphics[angle=0, scale=0.8]{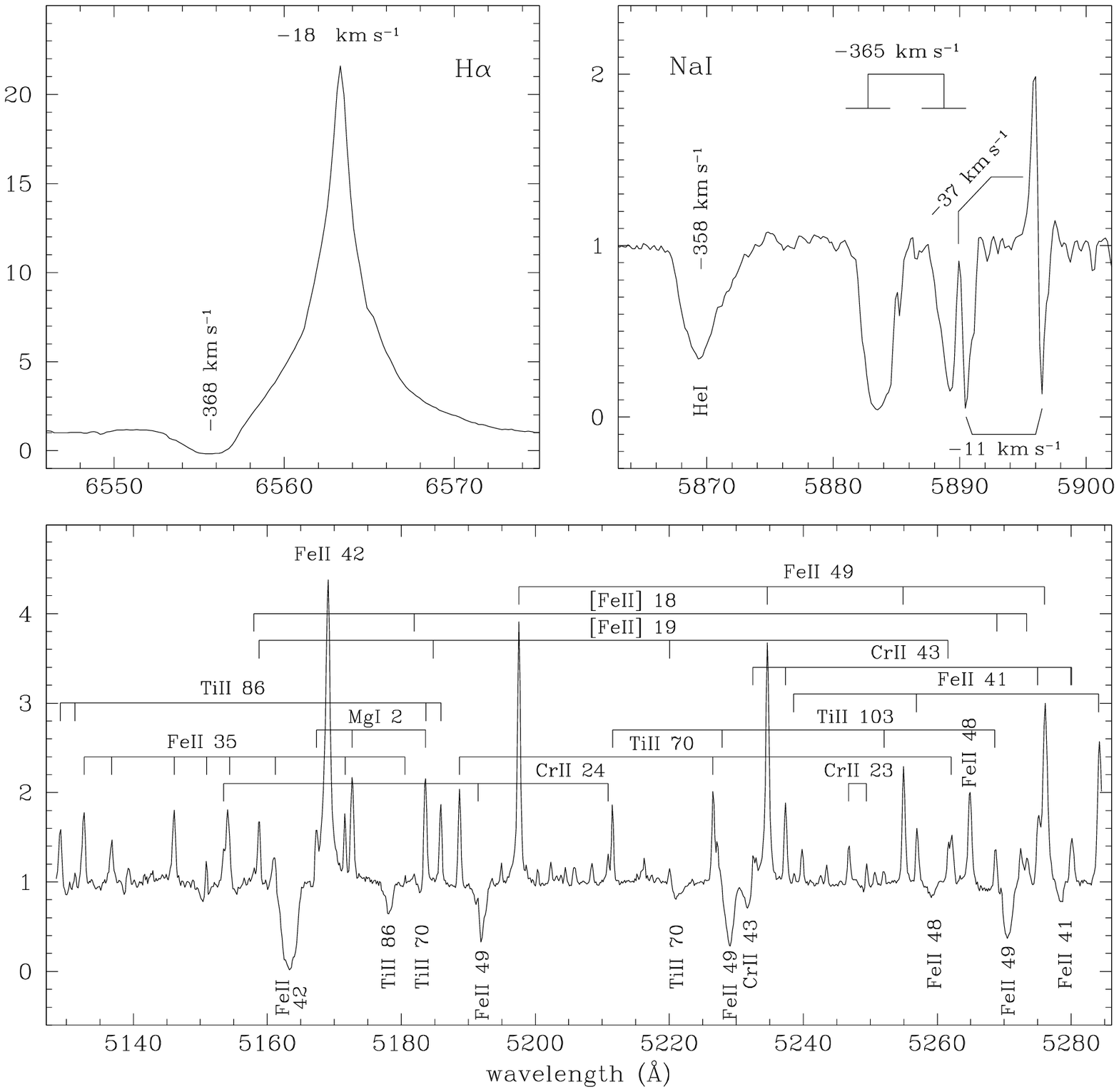}
\caption{As a guidance to the characteristics of the spectra included in this atlas,
some portions of the spectrum of XX~Oph are presented. The spectrum was
obtained on 7.90 July 2001 (UT) and the panels shows some of the most
popular emission lines: H$\alpha$, HeI/NaI, and the region of FeII multiplet 
49.
\label{fig1}}
\end{figure}

\clearpage
\begin{figure}
\includegraphics[angle=0, scale=0.8]{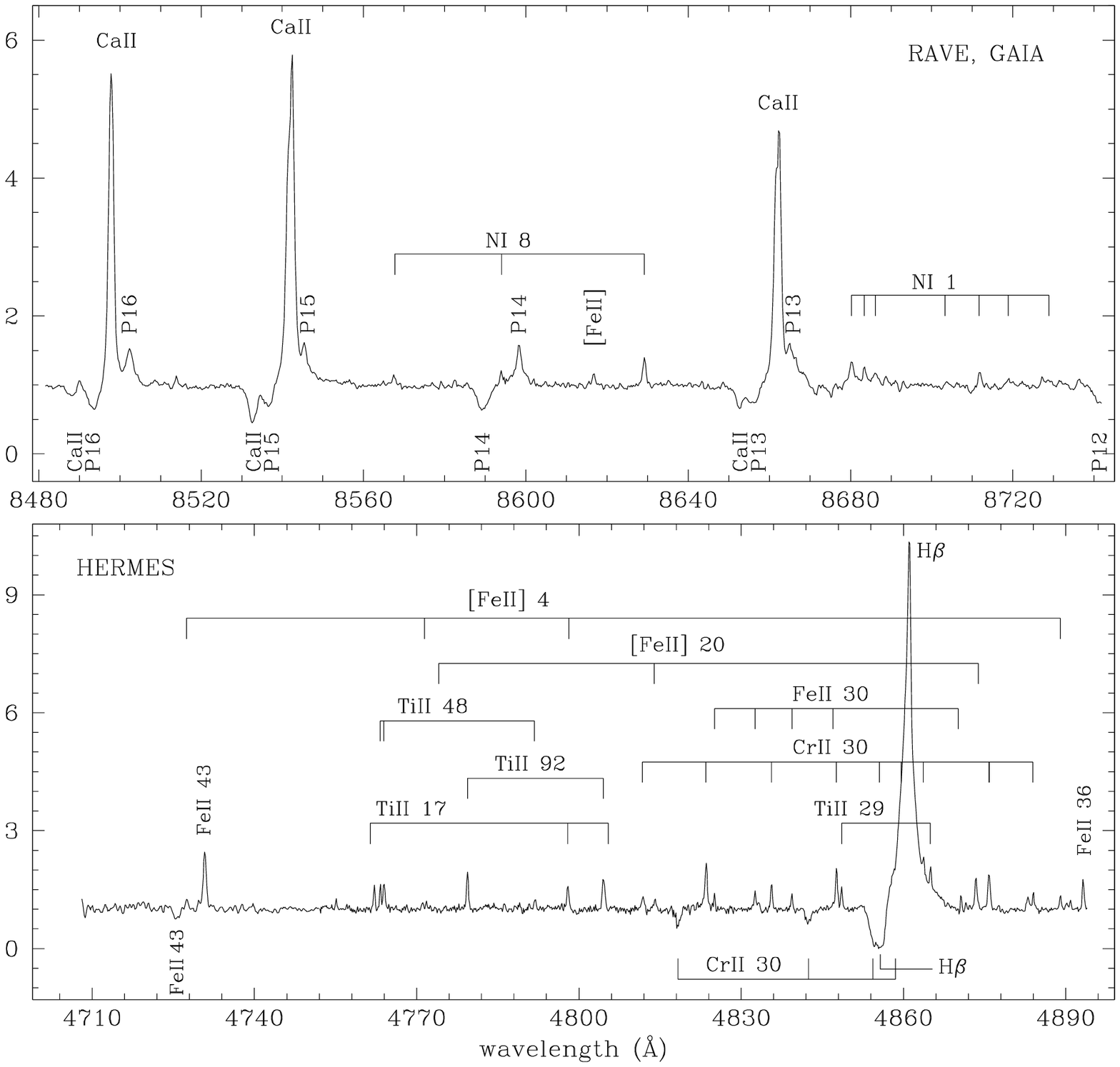}
\caption{The same spectrum of XX~Oph used for Figure~1, is presented here 
for the wavelength intervals covered by the RAVE survey and the GAIA space 
mission (upper panel), and the bluest of the four intervals observed by 
the HERMES survey (lower panel).
\label{fig1}}
\end{figure}

\clearpage

\begin{deluxetable}{l l l l l l c c }
\tabletypesize{\scriptsize}
\tablecaption{List of the program stars\label{tab1}}
\tablewidth{495.35pt}
\tablehead{
\colhead{Name~~~~~~~~} & \colhead{HD~~~~~~} & \colhead{Type~~~~~~~~~~} &
\colhead{GCVS~~~~~~~} &
\colhead{Spectrum~~~~~~~~} &
\colhead{Ref~~~~~~~~~} & \colhead{$(B-V)_T$} & \colhead{$V_T$}
}
\startdata
\cutinhead{\bf Chemically peculiar}
47 And        & 8374  & Am,Ap,SrCrEu       &          & A1m                &   AJ 74, 375  &    0.31   &   5.62 \\
NY Aur        & 51418 & Am,Ap,SrCrEu       &ACV	      & A0p(Eu-Sr-Cr)      &   GCVS        &    0.09   &   6.69 \\       
$\tau$ UMa    &78362  & Am,Ap,SrCrEu       &          & A . M,SB           &   ApJ 107, 109&    0.40   &   4.68 \\       
AF Dra        &196502 & Am,Ap,SrCrEu       &ACV	      & A 0 P SR CR EU     &   AJ 74, 375  &    0.11   &   5.20 \\       
$\beta$ CrB   &137909 & Am,Ap,SrCrEu       &ACV	      & F0p                &   ApJ 121, 653&    0.33   &   3.70 \\       
HD 178717     &178717 & Ba star            &          & Kp                 &   PASP 69, 326&    2.24   &   7.34 \\       
HD 199939     &199939 & Ba star            &          & G9 III:BaII        &   KFNT 17, 409&    1.54   &   7.58 \\       
DN CMi        &       & Carbon star        &SRB	      & C(Np)              &   GCVS    &\multicolumn{2}{r}{$\it~~~~(9.3)^a$}  \\       
RY Dra        &112559 & Carbon star        &SRB:      & C 7 I              &   ApJ 167, 521&    4.22   &   6.91 \\       
HD 198269     &198269 & CH star            &          & R0 .               &   AJ 63, 477  &    1.52   &   8.26 \\       
HR 6791       &166208 & CH star            &          & G8 III P:,SB:      &   ApJ 127, 172&    1.05   &   5.11 \\       
HR 6152       &148897 & CN star            &          & G8p                &   ApJ 116, 122&    1.48   &   5.38 \\       
91 Her        &163770 & CN star            &          & K1IIvar            &   ApJ 112, 362&    1.61   &   4.00 \\       
VX And        &1546   & J stars            &SRA	      & C 8                &   ApJ 167, 521&    5.74   &   8.37 \\       
DD Lyn        &64491  & $\lambda$ Boo      &DSCTC     & A3p                &   AJ 74, 375  &    0.29   &   6.26 \\       
HD 84123      &84123  & $\lambda$ Boo      &          & F0p                &   ApJ 113, 304&    0.32   &   6.88 \\       
$\lambda$ Boo &125162 & $\lambda$ Boo      &          & A0sh               &   AJ 74, 375  &    0.10   &   4.19 \\       
HD 105058     &105058 & $\lambda$ Boo      &          & A2p                &   AJ 73, 152  &    0.20   &   8.90 \\       
HR 3164       &66684  & $\lambda$ Boo      &          & B9.5Vp+A0Vp        &   GCVS        &\multicolumn{2}{r}{$\it~~~~(6.7)^a$}\\       
\cutinhead{\bf Pulsating}
$\beta$ Cep   &205021 & $\beta$ Cep        &BCEP      & B2 III SB,V        &   ApJS 2, 41  & $-$0.24~~   &   3.19 \\                         
$\beta$ CMa   &44743  & $\beta$ Cep        &BCEP      & B1II/III           &   MSS C04 0H  & $-$0.19~~   &   1.93 \\                         
BW Vul        &199140 & $\beta$ Cep        &BCEP      & B2IIIvar           &   ApJS 2, 41  & $-$0.17~~   &   6.53 \\                         
W Vir         &116802 & Ceph. pop.II       &CWA       & G6Ibvar            &   ApJS 2, 365 &    0.75   &   9.76 \\                         
AL Vir        &123984 & Ceph. pop.II       &CWA       & F3/F5III           &   MSS Vol.4   &    0.53   &   9.73 \\                         
AU Peg        &       & Ceph. pop.II       &CEP       & F8                 &   GCVS        &    1.01   &   9.27 \\                         
$\beta$ UMi   &131873 & Ceph. pop.II       &          & K4IIIvar           &   ApJ 116, 122&    1.78   &   2.24 \\                         
$\eta$ Aql    &187929 & Classical Ceph.    &DCEP      & F6 IB SB,V         &   ApJS 1, 175 &    0.87   &   3.96 \\
X Cyg         &197572 & Classical Ceph.    &DCEP      & G8Ib               &   ApJS 1, 175 &    1.49   &   6.62 \\
T Vul         &198726 & Classical Ceph.    &DCEP      & F5Ib               &   ApJ 131, 330&    0.66   &   5.67 \\
$\delta$ Del  &197461 & $\delta$ Sct       &DSCT      & DELTA DEL          &   AJ 74, 375  &    0.32   &   4.46 \\
28 Aql        &181333 & $\delta$ Sct       &DSCTC     & F0III              &   PASP 86, 70 &    0.28   &   5.56 \\
DQ Cep        &199908 & $\delta$ Sct       &DSCT      & F2II               &   PASP 64, 192&    0.40   &   7.29 \\
$\delta$ Sct  &172748 & $\delta$ Sct       &DSCT      & F 2 IIIP-DELDEL    &   PASP 86, 70 &    0.39   &   4.74 \\
$\tau$ Cyg    &202444 & $\delta$ Sct       &DSCT      & F1IV               &   AJ 81, 245  &    0.43   &   3.78 \\
$\chi$ Cyg    &187796 & Mira               &M         & S7.1 . E:          &   ApJ 120, 484&    1.83   &   6.96 \\
$o$ Cet       &14386  & Mira               &M         & M5e-M9e            &   GCVS        &    1.12   &   6.64 \\
R Cas         &224490 & Mira               &M         & M6e-M10e           &   GCVS        &    1.03   &   8.31 \\
S Cep         &206362 & Mira               &M         & C 6 II             &   ApJ 167, 521&    4.85   &   8.29 \\
X Oph         &172171 & Mira               &M         & K1 III COMP,V      &   ApJ 130, 611&    1.42   &   7.54 \\
RR Lyr        &182989 & RR Lyr             &RRAB      & A5.0-F7.0          &   GCVS        &    0.46   &   7.97 \\
SW And        &       & RR Lyr             &RRAB      & F8IIIvar           &   GCVS        &    0.47   &   9.75 \\
X Ari         &19510  & RR Lyr             &RRAB      & A8-F4              &   GCVS        &    0.50   &   9.68 \\
AC Her        &170756 & RV Tau             &RVA       & F4Ibpvar           &   ApJ 113, 60 &    0.74   &   7.70 \\
R Sct         &173819 & RV Tau             &RVA       & K0Ibpvar           &   ApJ 113, 60 &    1.52   &   5.54 \\
RV Tau        &283868 & RV Tau             &RVB       & K3pvar             &   ApJ 113, 60 & \multicolumn{2}{r}{$\it ~~~(9.8~13.3)^b$}\\
GZ Peg        &218634 & Semiregular        &SRA       & M4SIII             &   GCVS        &    1.62   &   5.29 \\
$\mu$ Cep     &206936 & Semiregular        &SRC       & M2Ia               &   ApJ 117, 313&    2.66   &   4.28 \\
$\rho$ Cas    &224014 & Semiregular        &SRD       & F8Iavar            &   PASP 69, 31 &    1.41   &   4.64 \\
R Lyr         &175865 & Semiregular        &SRB       & M5IIIvar           &   ApJ 101, 265&    1.64   &   4.34 \\
\cutinhead{\bf Interacting and outbursting}
68 Her        &156633 & $\beta$ Lyr        &EA/SD     & B1.5Vp             & ApJS 17, 371  & $-$0.19~~   &   4.78 \\                         
$\beta$ Lyr   &174638 & $\beta$ Lyr        &EB        & A8 :V COMP,SB      & PASP 72, 348  &    0.00   &   3.52 \\                         
CQ Dra        &108907 & CV                 &LB:       & M3IIIa             & GCVS          &    1.87   &   5.22 \\                         
SS Cyg        &206697 & CV                 &UGSS      & K5V+pec(UG)        & GCVS          & \multicolumn{2}{r}{$\it~~~(8.2-12.1)^b$}\\                         
WZ Sge        &       & CV                 &UGSU+E+ZZ & DAep(UG)           & GCVS          &\multicolumn{2}{r}{$\it ~~~(7.0-15.5)^b$}\\                         
AG Dra        &       & Symbiotic          &ZAND      & K1IIpevar          & ApJ 131, 83   &    1.71   &   9.97 \\                         
AX Per        &       & Symbiotic          &ZAND      & M3IIIep+A0         & GCVS          &\multicolumn{2}{r}{$\it ~~~(10.8-13.0)^b$}\\                         
Nova Cyg 2001 &       & Classical nova     &NA        &                    &               &         &         \\                         
V607 Aql      &       & Nova               &M         & M9                 & GCVS          &         &         \\                         
GK Per        &21629  & Old nova           &NA+XP     & pec(NOVA)          & GCVS          &\multicolumn{2}{r}{$\it ~~~(14.0)^b$}\\                         
T CrB         &143454 & Recurrent nova     &NR        & M3III+pec(NOVA)    & GCVS          &    1.57 &  10.29 \\                         
Cyg X-1       &226868 & high/m XRB         &ELL+XF    & B0Ib               & ApJS 2, 41    &    0.81 &   9.02 \\                         
V615 Cas      &       & high/m XRB         &*         & B1eIb:             & GCVS          &    0.70 &  10.90 \\                         
X Per         &24534  & high/m XRB         &GCAS+XP   & O9.5pe             & ApJS 17, 371  &    0.08 &   6.80 \\                         
40 Eri        &26965  & low/m XRB          &UV        & K1V                & ApJS 2, 195   &\multicolumn{2}{r}{$\it ~~~~(4.4)^a$}\\                         
MWC 560       &       & MWC 560 like       &*         & M6III D            & MNRAS 390, 377&\multicolumn{2}{r}{$\it ~~~~(9.7)^a$}\\                         
XX Oph        &161114 & MWC 560 like       &*         & Ape                & PASP 80, 197  &    1.01   &   9.04 \\                         
$\eta$ Gem    &42995  & VV Cep             &SRA+EA    & M3III              & ApJ 101, 265  &    1.87   &   3.48 \\                         
$\zeta$ Aur   &32068  & VV Cep             &EA/GS     & K4 II COMP         & ApJS 1, 175   &    1.36   &   3.89 \\                            
\cutinhead{\bf Active surfaces, fast rotating}
BY Dra        &234677 & BY Dra             &BY+UV     & K7Vvar             & MiARI 8, 1    &    1.41   &   8.33 \\                                  
HN Peg        &206860 & BY Dra             &BY        & G0V                & AJ 74, 916    &    0.64   &   6.02 \\                                  
CM Cam        &51066  & FK Com             &FKCOM     & G5 D               & Simbad        & $-$1.09~~   &   7.10 \\                                  
FK Com        &117555 & FK Com             &FKCOM     & G5II               & PDDO 2, 105   & $-$0.47~~   &   7.65 \\                                  
V645 Mon      &65953  & FK Com             &FKCOM     & K4III              & ApJ 116, 122  &    1.78   &   4.85 \\                                  
EV Lac        &       & Flare star         &UV+BY     & M4.5Ve             & GCVS          &    1.66   &  10.40 \\                                  
$\alpha^2$ CVn&112413 & Magnetic           &ACV       & A 0 P SI EU HG     & AJ 74, 375    &    0.06   &   2.87 \\                                  
V1264 Cyg     &184905 & Magnetic           &ACV       & A0p                & ApJ 128, 228  &    0.02   &   6.62 \\                                  
$\lambda$ And &222107 & RS CVn             &RS        & G8III-IV           & ApJ 117, 313  &    1.14   &   3.97 \\                                  
RS CVn        &114519 & RS CVn             &EA/AR/RS  & K2III              & ApJ 123, 246  &    0.66   &   8.29 \\                                  
\cutinhead{\bf Young emission-line objects}
AB Aur        &31293  & Herbig Ae/Be       &INA       & A0pe               & AJ 73, 588    &    0.14  &   7.08 \\                         
T Ori         &       & Herbig Ae/Be       &INSA      & A3V                & PASP 58, 366  & \multicolumn{2}{r}{$\it ~~~(12.6-9.5)^b$}\\                         
V380 Cep      &200775 & Herbig Ae/Be       &INA       & B2Ve               & PASP 80, 197  & $-$0.32~~   &   7.47 \\                         
V594 Cas      &       & Herbig Ae/Be       &INA       & B EQ               & AJ 73, 588    &    0.61   &  10.69 \\                         
V700 Mon      &259431 & Herbig Ae/Be       &INA       & B6pe               & ApJS 2, 41    &    0.26   &   8.83 \\                         
FU Ori        &       & pre-ZAMS           &FU        & G3Iavar            & GCVS          &\multicolumn{2}{r}{$\it ~~~(16.5-9.7)^b$}\\                         
$\nu$ Ori     &41753  & pre-ZAMS           &          & B3IV               & ApJS 17, 371  &    0.18   &   4.39 \\                         
RW Aur        &240764 & pre-ZAMS           &INT       & G5Ve(T)            & GCVS          &    0.48   &  10.38 \\                         
T Tau         &284419 & pre-ZAMS           &INT       & K0IIIe             & AJ 73, 588    &    1.31   &  10.01 \\                         
V863 Cas      & 4004  & Wolf Rayet         &WR        & WN 5 B             & Simbad        & $-$0.63~~   &  10.18 \\                         
WR 5          &17638  & Wolf Rayet         &          & WR D               & AJ 125, 2531  &    0.72   &  10.58 \\                         
Merrill's star&       & Wolf Rayet         &E:/WR     & WN8 .              & ApJ 96, 15    &\multicolumn{2}{r}{$\it ~~~~(11.2)^a$}\\                         
\cutinhead{\bf Other types}
HR 716        &15253  & shell star         &          & A2psh              &   AJ 74, 375  & $-$0.09~~   &    6.49 \\                         
48 Lib        &142983 & shell star         &GCAS      & B8Ia/Iab           &   MSS Vol.4   & $-$0.09~~   &    4.93 \\                         
P Cyg         &193237 & shell star         &SDOR      & B2pe               &   ApJS 17, 371&    0.40   &    4.83 \\                         
$\gamma$ Cas  &5394   & Be, $\gamma$ Cas   &GCAS      & B0IV:evar          &   ApJS 2, 41  &    0.06   &    2.17 \\                         
$\phi$ Per    &10516  & Be, $\gamma$ Cas   &GCAS      & B2Vpe              &   ApJS 17, 371&    0.11   &    4.02 \\                         
V568 Cyg      &197419 & Be, $\gamma$ Cas   &GCAS      & B2IV-Ve            &   ApJS 17, 371&    0.18   &    6.65 \\                         
V743 Mon      & 50138 & Be, $\gamma$ Cas   &GCAS      & B9 D               &   Simbad      &    0.03   &    6.58 \\                         
V1155 Tau     &32991  & Be                 &BE        & B2Ve               &   ApJS 17, 371&    0.18   &    5.86 \\                         
V725 Tau      &245770 & Be                 &XNGP      & Bpe                &   ApJS 2, 389 &    0.55   &    9.24 \\                         
Red Rectangle & 44179 & DIBs,UIPs,ISM      &R:        & B8V D              &   Simbad      &    0.36   &    9.06 \\                         
AE Aur        &34078  & O runaway          &INA       & O9.5Vvar           &   ApJS 2, 41  &    0.21   &    6.02 \\                
NGC 7027      &201272 & PN                 &          & Pe PN              &   KFNT 17, 409&\multicolumn{2}{r}{$\it ~~~~(9.7)^a$}     \\                
HD 8550       &8550   & post-AGB, pre-PN   &          & F0 D               &   Simbad      &\multicolumn{2}{r}{$\it ~~~~(9.6)^c$} \\                
UU Her        &       & post-AGB, pre-PN   &SRD       & F8VIIvar           &   ApJ 112, 554&    0.57   &    9.12 \\                
R CrB         &141527 & R CrB              &RCB       & C0,0(F8pep)        &   GCVS        &    0.65   &    5.96 \\                
XX Cam        &25878  & R CrB              &RCB:      & G1I(C0-2,0)        &   GCVS        &    0.93   &    7.39 \\                
HD 29537      &29537  & $v_{rot}$          &          & F0                 &   Simbad      &    0.42   &    6.82 \\                         
NGC 3031      &       & galactic bulge     &          &                    &               &           &         \\  
\enddata
\tablenotetext{a}{$V_J$ from VizieR: catalogue I/280B/ascc, (Kharchenko+
2009)}
\tablenotetext{b}{$V_{max}$ and $V_{min}$ from GCVS: catalogue B/gcvs,
(Samus+ 2007-2010)}
\tablenotetext{c}{Mean $V$ magnitude from VizieR: catalogue II/215 (Hauck+
1997)}
\end{deluxetable}

\clearpage

\begin{deluxetable}{l c c c c c c c c c c c c}
\tabletypesize{\scriptsize}
\rotate
\tablecaption{Atlas products\label{tab2}}
\tablewidth{0pt}
\tablehead{
\colhead{Fits file} & \colhead{HJD} & \colhead{Exp.} &
\colhead{Star} &
\colhead{HD} &
\colhead{Peculiarity} & \colhead{Spec. type} & \colhead{$\Delta\lambda$}
&\colhead{orders} & \colhead{S/N} & \colhead{std} &
\colhead{tell.} &\colhead{notes}\\
\colhead{~~} & \colhead{~~} & \colhead{(sec)} &
\colhead{~~} &
\colhead{~~} &
\colhead{~~} & \colhead{~~} & \colhead{~~}
&\colhead{$\#$} & \colhead{($\#$34)} & \colhead{~~} &
\colhead{~~} &\colhead{~~}

}
\startdata
47And\_1&2452125.628 &   720  & 47 And    & 8374 &Am,Ap,SrCrEu            &   A1m   &4600-9450&48-24& 200&1&1,2&\\
47And\_2&2452158.563 &   900  &           &      &                        &         &4600-9450&48-24& 230&&3&\\
47And\_3&2452158.577 &   900  &           &      &                        &         &4600-9450&48-24& 230&&3&\\
47And\_4&2452158.590 &   900  &           &      &                        &         &4600-9450&48-24& 230&&3&\\
\tableline
NYAur\_1&2452359.374 &   900  & NY Aur    & 51418&Am,Ap,SrCrEu       &A0p(Eu-Sr-Cr) &4575-9450&48-24& 100&2,3&4,5,6,7&Przybylski star\\
NYAur\_2&2452359.390 &   900  &           &      &                   &              &4575-9450&48-24& 120&2,3&4,5,6,7&\\
\tableline
tauUMa\_1&2451199.611&   90   & $\tau$ UMa&78362 &Am,Ap,SrCrEu            & A . M,SB&4600-9300&48-24&  80&&&\\
tauUMa\_2&2451199.615&   90   &           &      &                        &         &4600-9300&48-24&  80&&&\\
tauUMa\_3&2451199.618&   90   &           &      &                        &         &4600-9300&48-24&  80&&&\\
\tableline
AFDra\_1 &2452363.582&  120   & AF Dra    &196502&Am,Ap,SrCrEu        &A0p(Eu-Sr-Cr)&4600-9450&48-24& 160&2,3&4,5,6,7&\\
AFDra\_2 &2452363.586&  120   &           &      &                     &            &4600-9450&48-24& 160&2,3&4,5,6,7&\\
AFDra\_3 &2452363.589&  120   &           &      &                     &            &4600-9450&48-24& 160&2,3&4,5,6,7&\\
AFDra\_4 &2452363.593&  120   &           &      &                     &            &4600-9450&48-24& 160&2,3&4,5,6,7&\\
AFDra\_5 &2452363.597&  120   &           &      &                     &            &4600-9450&48-24& 160&2,3&4,5,6,7&\\
\enddata
\tablecomments{Table~2 is published in its entirety in the 
electronic edition of the {\it Astronomical Journal}.  A portion is 
shown here for guidance regarding its form and content.}
\end{deluxetable}

\clearpage

\begin{deluxetable}{l l l l r l}
\tabletypesize{\scriptsize}
\tablecaption{Spectrophotometric standard stars\label{tab3}}
\tablewidth{0pt}
\tablehead{
\colhead{std}  &
\colhead{Fits file} &
\colhead{Star} &
\colhead{HJD } & 
\colhead{Exp.(sec)} & 
\colhead{Airmass}
}
\startdata
1             & HR7596\_1  & HR 7596   &  2452125.393 & 300 & 1.49\\
2             & HR5511\_2  & HR 5511   &  2452361.519 & 80  & 1.47\\
3             & HR5511\_3  & HR 5511   &  2452362.571 & 40  & 1.39\\
4             & HR5511\_4  & HR 5511   &  2452062.368 & 120 &  1.41\\
5             & HR3454\_5  & HR 3454   &  2415046.480 & 110 &  1.37 \\
6             & HR4468\_6  & HR 4468   &  2415070.520 & 210 &  1.84 \\
\enddata
\end{deluxetable}

\clearpage

\begin{deluxetable}{l l l l r l}
\tabletypesize{\scriptsize}
\tablecaption{Telluric dividers\label{tab4}}
\tablewidth{0pt}
\tablehead{
\colhead{tell.}  &
\colhead{Fits file} &
\colhead{Star} &
\colhead{HJD } & 
\colhead{Exp.(sec)} & 
\colhead{Airmass}
}
\startdata
1             & HR5778\_1  & HR 5778   &  2452123.323 & 120 & 1.10\\
2             & HR7040\_2  & HR 7040   &  2452125.361 & 360 & 1.41\\
3             & HR2209\_3  & HR 2209   &  2452158.610 & 840 & 1.25\\
4             & HR2659\_4  & HR 2659   &  2452359.267 & 300 & 1.09\\
5             & HR5778\_5  & HR 5778   &  2452363.640 &  30 & 1.06\\
6             & HR5511\_6  & HR 5511   &  2452361.519 & 80  & 1.47\\
7             & HR5511\_7  & HR 5511   &  2452362.571 & 40  & 1.39\\
8             & HR5511\_8  & HR 5511   &  2452062.368 & 120 &  1.41\\
9             & HR809\_9   & HR 809    &  2452212.500 & 600 & 1.17\\
10            & HR1910\_10  & HR 1910   &  2452213.653 & 20  & 1.15\\
11            & HR2648\_11  & HR 2648   &  2452216.710 & 650 & 1.74\\
12            & HR6175\_12  & HR 6175   &  2451258.663 & 14  & 1.84\\
13            & HR7040\_13  & HR 7040   &  2452097.401 & 600 &  1.52\\
14            & HR5778\_14  & HR 5778   &  2452098.348 & 180 &  1.04\\ 
15            & HR6175\_15  & HR 6175   &  2452506.320 & 260 &   2.06\\
16            & HR7894\_16  & HR 7894   &  2452447.525& 450 &   1.12\\
17            & HR496\_17   & HR 496    &  2451542.260 & 80 &    1.00\\
18            & HR496\_18   & HR 496    &   2415046.211&140 &    1.01\\
19            & HR2209\_19  & HR 2209   &   2415070.501 & 150 &  1.30\\
\enddata
\end{deluxetable}

\clearpage

\begin{deluxetable}{c c c}
\tabletypesize{\scriptsize}
\tablecaption{Mean heliocentric velocities (km sec$^{-1}$) for the emission and absorption lines 
of ions in the spectrum of XX Oph presented in Figures 1 and 2\label{tab5}}
\tablewidth{0pt}
\tablenum{5}
\tablehead{
\colhead{lines} & \colhead{emission component} & \colhead{absorption
component} 
}
\startdata
            H$\alpha$, H$\beta$     &   $-$17& $-$370\\
            Paschen    &   $-$37& $-$354\\
            CaII       &   $-$38& $-$367\\
            CrII       &   $-$37& $-$367\\
            NaI        &   $-$37& $-$366\\
            TiII       &   $-$38& $-$360\\
            FeII       &   $-$37& $-$361\\
            $\lbrack$FeII$\rbrack$   &   $-$37&  \\
            MgI        &   $-$38&  \\
            NI         &   $-$36&  \\
\enddata
\end{deluxetable}

\end{document}